\documentclass{emulateapj}

\usepackage{color} 

\slugcomment{}

\shorttitle{}
\shortauthors{Koshida et al.}

\begin{document}

\title{Calibration of AGN reverberation distance measurements}

\author{Shintaro Koshida\altaffilmark{1}, Yuzuru Yoshii\altaffilmark{2,3,}\footnotemark[8], Yukiyasu Kobayashi\altaffilmark{4}, Takeo Minezaki\altaffilmark{2}, Keigo Enya\altaffilmark{5}, Masahiro Suganuma\altaffilmark{4}, Hiroyuki Tomita\altaffilmark{2}, Tsutomu Aoki\altaffilmark{6}, and Bruce A. Peterson\altaffilmark{7}}

\altaffiltext{1}{Subaru Telescope, National Astronomical Observatory of Japan, 650 North A\'ohoku Place, Hilo, Hawaii, 96720, USA}
\altaffiltext{2}{Institute of Astronomy, School of Science, University of Tokyo, 2-21-1 Osawa, Maka, Tokyo 181-0015, Japan}
\altaffiltext{3}{Steward Observatory, University of Arizona, 933 North Cherry Avenue, Rm. N204, Tucson, AZ 85721-0065, USA}
\altaffiltext{4}{National Astronomical Observatory of Japan, 2-21-1 Osawa, Mitaka, Tokyo 181-8588, Japan}
\altaffiltext{5}{Institute of Space and Astronomical Science, Japan Aerospace Exploration Agency, 3-1-1, Yoshinodai, Sagamihara, Kanagawa 229-8510, Japan}
\altaffiltext{6}{Kiso Observatory, Institute of Astronomy, School of Science, University of Tokyo, 10762-30 Mitake, Kiso, Nagano 397-0101, Japan}
\altaffiltext{7}{Mount Stromlo Observatory, Research School of Astronomy and Astrophysics, Australian National University, Weston Creek P.O., ACT 2611, Australia}
\footnotetext[8]{The principal investigator of the MAGNUM project}

\begin{abstract}
In \citet{yosh14}, we described a new method for measuring extragalactic distances based on dust reverberation in active galactic nuclei (AGNs), and we validated our new method with Cepheid variable stars. In this paper, we validate our new method with Type Ia supernovae (SNe Ia) which occurred in two of the AGN host galaxies during our AGN monitoring program: SN 2004bd in NGC 3786 and SN 2008ec in NGC 7469. Their multicolor light curves were observed and analyzed using two widely accepted methods for measuring SN distances, and the distance moduli  derived are $\mu=33.47\pm 0.15$ for SN 2004bd and $33.83\pm 0.07$ for SN 2008ec. These results are used to obtain independently the distance measurement calibration factor, $g$.
The $g$ value obtained from the SN Ia discussed in this paper is $g_{\rm SN} = 10.61\pm 0.50$ which matches, within the range of 1$\sigma$ uncertainty, $g_{\rm DUST} = 10.60$, previously calculated {\it ab initio} in \citet{yosh14}.
Having validated our new method for measuring extragalactic distances, we use our new method to
calibrate reverberation distances derived from variations of H$\beta$ emission in the AGN broad line region (BLR), extending the Hubble diagram to $z\approx 0.3$ where distinguishing between cosmologies is becoming possible.
\end{abstract}

\section{Introduction}

An AGN consists of a hot central engine, thought to be an accretion disk around a black hole, a dust free region of ionized gas surrounding the accretion disk where broad emission lines originate, and a dust torus that lies beyond the BLR. In \citet{yosh14}, we used the spectral energy distribution (SED) of the central engine and the dust properties to calculate the light travel time from the central engine to the dust torus in terms of the absolute luminosity of the central engine. We measured the light travel times, 
$\Delta t$ in days, for 17 AGNs, and derived luminosity distances, $d$, in Mpc, using
\begin{equation}
d=\Delta t\times 10^{0.2(m_{V}-A_{V}-k_{V}-25+g)} \label{eq1}
\end{equation}
where $m_{V}$ is the mean observed apparent magnitude of the AGN in the $V$ band, $A_{V}$ is the Galactic extinction in the $V$ band, $k_{V}$ is the $K$-correction for the $V$ band, and $g=g_{\rm DUST}=10.60$ is the calculated calibration factor that characterizes the dust sublimation by the radiation field of the central engine \citep[for details, see][]{yosh14}.  We obtained $H_0=73\pm 3$ km s$^{-1}$ Mpc$^{-1}$ in good agreement with $H_0=75\pm 10$ km s$^{-1}$ Mpc$^{-1}$ obtained from empirically calibrated Cepheid variable stars \citep{free01,free02}.

During our AGN monitoring program for the MAGNUM project \citep{yosh02,yosh03}, SNe Ia occurred in two of the monitored AGN host galaxies, SN 2004bd in NGC 3786, and SN 2008ec in NGC 7469.  
In \S\ref{sec:SN}, we use our multicolor observations of the SN Ia to derive a luminosity distance, $d_{\rm SN}$, for each SN Ia, which we use in Equation (\ref{eq1}) to obtain $g_{\rm SN}$. In \S\ref{sec:DD} we compare $g_{\rm SN}$ with $g_{\rm DUST}$, and also discuss the error budget for the calculation of $g_{\rm DUST}$.

Fourteen of the AGNs studied in \citet{yosh14} are common in the study by \citet{wats11} to obtain distances from the delay between continuum flaring and an increase in the H$\beta$ emission line flux. 
They calibrated their relative distances using the surface brightness fluctuation distance of the companion galaxy of NGC 3227, $\mu =31.86\pm 0.24$. In \S\ref{sec:HB}, we calibrate their H$\beta$ delays using our distances in \citet{yosh14} for the AGNs that we have in common, and we derive $g_{{\rm H}\beta}$. We present our conclusions in \S\ref{sec:CON}.

\section{Supernovae}\label{sec:SN}

\subsection{Observations}\label{sec:obs}

Photometric data for SN 2004bd and SN 2008ec were obtained during the MAGNUM AGN monitoring program. The Multiwavelength Imaging Photometer \citep[MIP, ][]{koba98} on the MAGNUM telescope had a 90 arcsec field of view and obtained images including each host galaxy and its SN.
Observations were made in four optical photometric bands $U$, $B$, $V$, and $I$, from MJD 53,097 to 53,208 for SN 2004bd, and from MJD 54,661 to 54,76 for SN 2008ec.  For each SN, the basic characteristics are listed in Table \ref{tab:tab1}.

\subsection{Data Reduction}\label{sec:dr}

Standard reduction procedures were applied to the images, such as bias subtraction and flat fielding, as in \citet{kosh14, suga06, mine04}. Before proceeding to the photometry, the host galaxy flux was subtracted from each image using template galaxy images made as follows. First, we selected the images observed on nights just before the occurance of SN, which had seeing of one arcsec or better. For SN 2004bd, images of NGC 3786 obtained at MJD 53046.4 were used for all bands. For SN 2008ec, the images obtained at MJD 54649.6, 54651.6, 54653.5, and 54651.6 were used for the $U$, $B$, $V$, and $I$ bands, respectively. Second, we stacked all the images in a band, obtained on the night, in order to improve the signal to noise ratio of the template galaxy image.  Any variable AGN components in the template image should not affect the SN photometry, unless the position of the AGN is inside the photometric aperture.  However, SN 2004bd was only 5.3 arcsec from the nucleus, which would contaminate the photometry aperture and sky flux measurement area. To avoid AGN effects on the photometric results, we subtracted the AGN component from the template image using GALFIT \citep{peng02}, which models the surface brightness distribution of galaxies or point sources. The template image was subtracted from each observed image after adjusting the seeing size of the template image by Gaussian convolution using the IRAF task GAUSS. The varying AGN component in each observed image remained after the subtraction, so we subtracted it from each observed image using the GALFIT PSF model. 

We performed aperture photometry with an $8.3$ arcsec diameter aperture. Flux in the aperture was calibrated using standard stars listed in \citet{land92} that were observed on the same night.  The resultant light curves in the $U$, $B$, $V$, and $I$ band of SN 2004bd and SN 2008ec are plotted with filled circles in the upper and lower panels, respectively, in Figure \ref{fig:fig1}.
The photometry following the process above achieved a photometric error around $1$\% in flux.
\footnote{The Open Supernova Catalog project (https://sne.space/, arXiv:1605.01054) accumulates the photometric data of SNe in literature, including SN 2004bd and SN 2008ec, and provides their multicolor light curves. However, we used only the data of our own in this paper, in order to avoid possible biases which may arise from using heterogeneous data sets obtained with different instruments, photometric methods, host galaxy flux estimation methods, etc.} 

\subsection{Light Curve Fitting}

A variety of methods have been developed to correct the intrinsic dispersion of the peak brightness of SN Ia using light curve shapes \citep[e.g.][]{ries96,perl99}.  We selected two methods: the Multicolor Light Curve Shape method \citep[MLCS2k2,][]{jha07} and the second version of the Spectral Adaptive Light curve Template method \citep[SALT2,][]{guy07, guy10}, to measure the distances of the SNe Ia. These standard and sophisticated methods provide the template light or color index curves and the necessary basic parameters to model the multicolor light curves of any SN Ia, and measure its distance.

\subsubsection{MLCS2k2}

In the MLCS2k2 method, the observed light curve set of any SN Ia in $U$, $B$, $V$, $R$, and $I$ is fit using the parameter $\Delta$ with the template set trained by a sample of nearby SN Ia. The difference from the template curve at the peak magnitude in the $V$ band is represented by $\Delta$.

Using the color variation curve, the MLCS2k2 method can also estimate total extinction $A_{V}$ along the light path to the SN in addition to the peak magnitude. The wavelength independent distance modulus, $\mu$, can be obtained 
as $\mu=\mu_{\lambda}-A_{\lambda}$.
For instance, the model formula for the $V$ band is given as
\begin{eqnarray}
\mu &= m_{\lambda}(t_{0})-M_{\lambda}^{\text{peak}}-\zeta_{\lambda}\left(\alpha_{\lambda}+\frac{\beta_{\lambda}}{R_{V}}\right)A_{V}\nonumber \\ 
&-P_{\lambda}\Delta-Q_{\lambda}\Delta^{2},
\end{eqnarray}
where $A_{V}$ is total extinction in the $V$ band, and $\Delta$ is differential magnitude from template light curve at the peak in $V$, which parameterizes the whole light curve model with the coefficients $P_{\lambda}$ and $Q_{\lambda}$.  Details of parameters $\zeta_{\lambda}, \alpha_{\lambda},$ and $\beta_{\lambda}$, concerning the extinction estimation, are discussed in \citet{jha07}.

The correction for the Galactic extinction $A_{V,{\rm MW}}=0.079$ 
for SN 2004bd and $A_{V,{\rm MW}}=0.228$ for SN 2008ec, according 
to \citet{schl98}, was applied to the observed light curves prior 
fitting. The fit converged with a slightly better chi-square value 
than without the correction. 
We fixed $R_{V}=3.1$ following the discussion in \citet{jha07}. 
Most $R_{V}$ derived by MLCS2k2 are distributed the range $3.0\pm 0.1$.
Setting $R_{V}$ as a free parameter made our fitting unstable.

\begin{figure}
\vspace{12cm}
\includegraphics[width=1.5cm, bb=0 0 100 100]{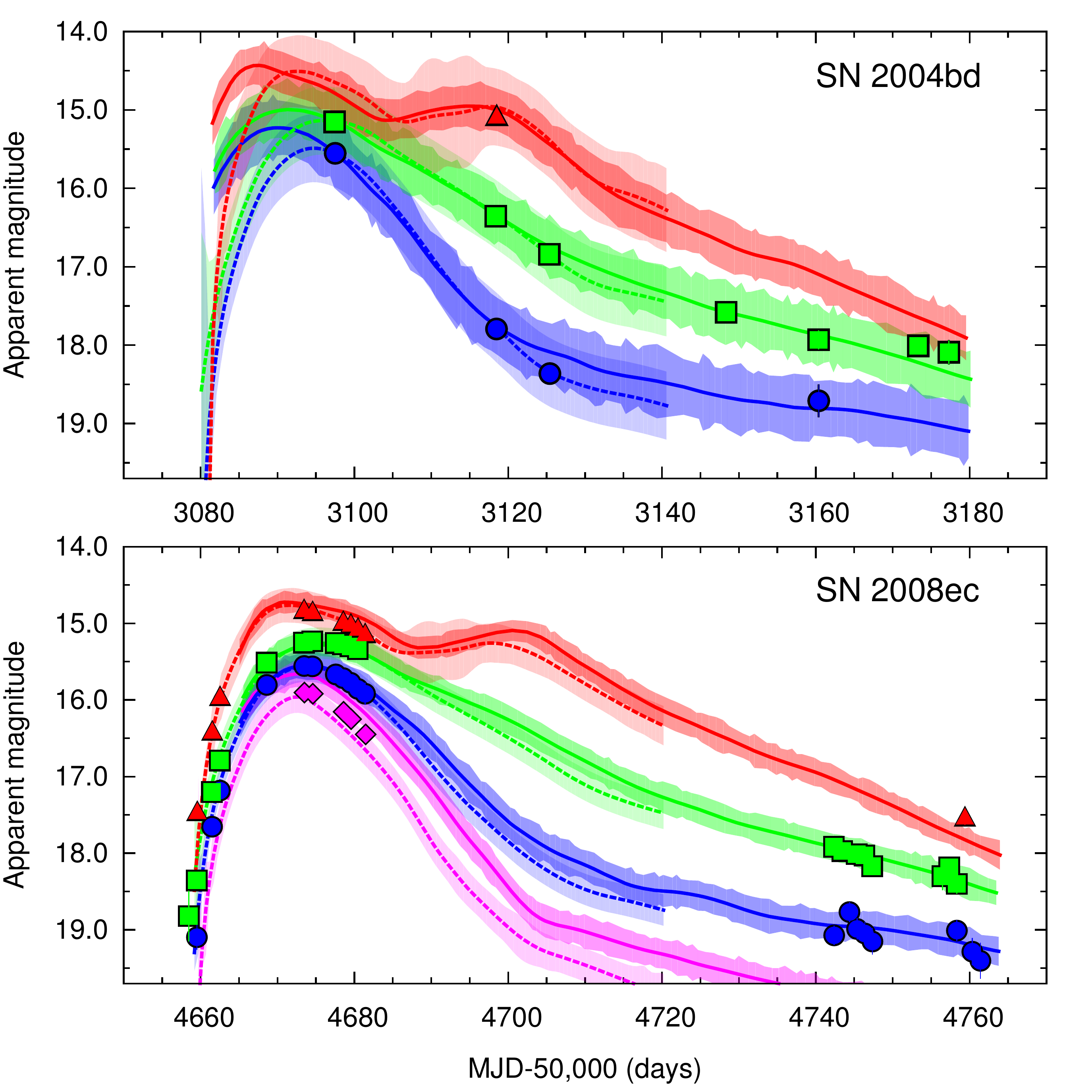}
\caption{Light curves of SN 2004bd in the upper panel and those of SN 2008ec in the lower panel. The MAGNUM observations are plotted as filled diamonds, circles, squares and triangles to represent $U+0.4$, $B$, $V$, and $I-0.4$ in apparent magnitude, respectively. The solid lines are the model light curves obtained using the MLCS2k2 method, while the dashed lines are the model light curves obtained with SALT2. Dark and light shadows along the curves indicate the error of the model light curves for MLCS2k2 and SALT2, respectively. }
\label{fig:fig1}
\end{figure}

The best fit model curves for SN 2004bd are shown in the upper panel of Figure \ref{fig:fig1} as solid lines. The derived distance modulus is $\mu=33.53\pm 0.24$ with a total extinction of $A_V=0.93\pm 0.29$ mag in the $V$ band. Similarly, the fitted model curves for the SN 2008ec observations are displayed in the lower panel of Figure \ref{fig:fig1}. The derived distance modulus is $\mu=33.88\pm 0.13$ with a total extinction of $A_V=1.03\pm 0.09$ in the $V$ band. The systematic errors in these distance moduli are $\pm 0.21$ for SN 2004bd and $\pm 0.18$ for SN 2008ec, originating from several error sources such as uncertainty of model parameters, intrinsic dispersion of SN Ia luminosity, and residuals of the training set light curves from the model curves. These results are listed in the Table \ref{tab:tab2}.

\subsubsection{SALT2}

SALT2 is a set of template light curves of SN Ia based on a training set from the spectroscopic monitoring of SN Ia \citep{guy07,guy10}.  They modeled the light curve set as
\begin{equation}
\mu=m^{*}_{\lambda}-M_{\lambda}+\alpha_{x} x_{1} -\beta c,\nonumber \label{eq:salt2mod}
\end{equation}
where $m^{*}_{\lambda}$ is a rest-frame magnitude at wavelength
 $\lambda$, $\alpha_{x}$ is the slope of light-curve 
width to luminosity relationship which encodes time-variable color as 
a function of light-curve shape, and $\beta$ is the 
coefficient for constant dust extinction. 
The values of the parameters obtained 
from the training set are reported in \citet{guy10}, and we adopt 
their results.

The SALT2 team provides their fitting code named SNFIT platformed 
on Python. We used their code, setting the range of wavelength and 
period to $2000${\AA }$<\lambda<9200${\AA } and $-20<t<50$ days, as 
similar as possible to the MLCS2k2 method. We again subtract the 
Galactic dust extinction from the light curves used for the fit, so 
the fitting model gives only the extinction in the SN host galaxy, 
as for MLCS2k2.

The best fit model light curves for SN 2004bd and SN 2008ec from 
SALT2 are overdrawn with dashed lines in the upper and lower panels, 
respectively, in Figure \ref{fig:fig1}.  The distance moduli obtained 
from this method are, for SN 2004bd, $\mu=33.43\pm 0.19$ with 
$A_V=0.82\pm 0.13$, and for SN 2008ec, $\mu=33.79\pm 0.08$ with 
$A_V=0.65\pm 0.07$.
Systematic errors in these distance moduli are $\pm 0.17$ for 
SN 2004bd and $\pm 0.16$ for SN 2008ec, considering the same error 
sources as in MLCS2k2.   

Our derived apparent magnitude at maximum 
luminosity in the B-band for SN 2008ec
is $m_{B, max}=15.51\pm 0.02$.
This is consistent with $15.49\pm 0.03$ in \citet{gane10} and 
 $15.51\pm 0.03$ in \citet{gane13}. Our derived color
parameter of $0.24\pm 0.03$ is consitent with $0.21\pm 0.03$ in
\citet{scal14}, but not with $0.09\pm 0.03$
in \citet{gane13}. We could not find any comparable results 
for SN 2004bd in the literature.

\subsection{SN Ia Distances}\label{sec:SNdis}

For each of the two SNe Ia, the distance moduli obtained from the 
different methods, MLCS2k2 and SALT2, are consistent. 
In order to insure that our relatively poor sampling of the SN 2004bd
light curve has not affected the fitting results, we included the
photometric data from \citet{gane10} and obtained $\mu=33.43\pm 0.16$
using MLCS2k2 and $\mu=33.37\pm 0.14$ using SALT2, which are consistent
with $\mu=33.43\pm 0.19$ with only our own data.
The light curve width parameter and the extinction are also
consistent within $1\ \sigma$. Therefore, we adopt the results from
our homogeneous data for further discussion.

With the standard dust extinction law, $R_{V}=3.1$ and therefore $A_{B}/A_{V}=1.34$ \citep{card89}, the total extinction in the $V$ band, $A_V$, estimated for SN 2004bd by the two methods is consistent within the large errors, while the two values of $A_V$ estimated for SN 2008ec differ by 0.4 mag.
This difference might come from including the $U$ band in the color determination for SN 2008ec, because, in both methods, the $U$ band training set uncertainties are large.

In order to avoid underestimating the errors, we adopt the simple mean
for $\mu$ and its error instead of the weighted mean, because the two methods are not independent. They use training sets with many data sets in common.
For SN 2004bd, $\mu=33.48\pm 0.22$, which corresponds to a luminosity 
distance of $d_{\rm SN}=49.68\pm 4.92$ Mpc. The systematic errors 
are $\pm 0.19$ in $\mu$ and $\pm 4.37$ Mpc in $d_{\rm SN}$.
For SN 2008ec, $\mu=33.84\pm 0.10$, corresponding to 
$d_{\rm SN}=58.48\pm 2.80$ Mpc. The systematic errors are 
$\pm 0.17$ in $\mu$ and $\pm 4.63$ Mpc in $d_{\rm SN}$.
 
Our SN Ia derived distances are consistent with those in the 
literature for the parent galaxies derived by other methods,
such as the Tully-Fisher relation \citep{scho94,theu07}, AGN 
reverberation \citep{coll99,cack07,wang14}, and galaxy ring structure 
diameter \citep{pedr81}. The mean and standard deviation of the 
results from these studies are $\mu=33.32\pm 0.15$ for NGC3786, and $\mu=34.02\pm 0.56$ for NGC7469.

\section{Dust Properties of AGNs}\label{sec:DD}

Substituting $d_{\rm SN}$ into the LHS of Equation (\ref{eq1}), we obtain the calibration factor of $g_{\rm SN} = 9.73 \pm 1.16$ for SN 2004bd and $g_{\rm SN} = 11.04 \pm 0.56$ for SN 2008ec. Taking both random errors and systematic errors into account, the weighted average for our two SNe Ia is $g_{\rm SN} = 10.61 \pm 0.50$, and the systematic error in $g_{\rm SN}$, from that in  $d_{\rm SN}$, is $\pm 0.69$. 
The calculated calibration factor $g_{\rm DUST}=10.60$ used in \citet{yosh14} is within 1$\sigma$ of $g_{\rm SN}$. This agreement provides an independent confirmation of the validity of our new method for measuring extragalactic distances described in \citet{yosh14}.

The standard deviation $\sigma_{g_{\rm SN}}$ for our two SNe Ia is 0.62, 
which, in principle, should be the combination of the random and 
systematic errors in $g_{\rm SN}$. 
The errors discussed above give the combined error of $\sqrt{0.50^{2}+0.69^{2}}=0.85$, which is greater than $\sigma_{g_{\rm SN}}$ indicating that the individual random and systematic errors are overestimated by taking the simple mean of the MLCS2k2 and SALT2 methods. Using the weighted mean, and ignoring any covariance from using similar training sets, gives random and systematic errors of $0.34$ and $0.49$ respectively, and a combined error of $0.60$, similar to $\sigma_{g_{\rm SN}}$.

An estimate of the cosmic variation in our dust-reverberation distances can be obtained by examining the differences between $d_{\rm DUST}$, as given by Equation (\ref{eq1}) with $g_{\rm DUST}=10.60$, and $d_H =v/H_0$ ($H_0=73\pm 3$ km s$^{-1}$ Mpc$^{-1}$) for the 17 AGNs studied in \citet{yosh14}.  In practice, by substituting $d_H$ into the LHS of Equation (\ref{eq1}), we express the cosmic variation of the dust properties in terms of the distribution of $g_{\rm DUST}$, which reflects random errors as well as systematic errors, including possible object to object variations.
Our dust reverberation model has three parameters: 
(1) the dust sublimation temperature, $T_{\rm d}=1700\pm 50$ K, derived from the $H-K$, $J-K$, and $J-H$ color temperatures of the variable near-infrared component in our sample of Seyfert 1 galaxies \citep{tomi06}, 
(2) the power-law index, $\alpha_{\rm UV} =-0.5 \pm 0.2$, for the UV to optical continuum emission from the central engine which heats the dust, as found in the SDSS sample of QSOs \citep{vand01,dave07}, and 
(3) the grain size, $a$, represented by the distribution of $f(a)\propto a^{-2.75}$ ($0.005\ \mu{\rm m} \leq a \leq 0.20\ \mu{\rm m}$) that is intermediate between the two opposite extreme distributions of grain size in the literature \citep[see ][]{yosh14}.

Using $d_H$ in the LHS of Equation (\ref{eq1}), the standard deviation of $g_{\rm DUST}$ for the 17 AGNs studied in \citet{yosh14} is derived as $0.44$. This must include the well estimated errors of $\pm 0.24$ from our $\Delta t_{\rm DUST}$ measurements, $\pm 0.1$ from intrinsic extinction \citep{cack07}, $\pm 0.18$ from the possible range of $T_{\rm d}$, and $\pm 0.32$ from $\alpha$. 
The remaining, necessary error is $\pm 0.09$ and is attributed to the systematic error from the grain size for $f(a)$, much less than the conceivable maximum error $\sim\pm 0.5$ \citep{yosh14} from assuming the two opposite extreme distributions of $a$.  Thus, the dust properties of all AGNs are not very different. 

\citet{hoen17} claimed that distances measured by \citet{yosh14} suffered from the degeneracy between $H_{0}$ and dust emissivity, because AGN luminosity was calculated by redshift-based distance. This claim is not correct because \citet{yosh14} calculated the luminosity based on the physical model of radiative thermal equilibrium \citep[e.g.][]{barv87} which is independent of $H_{0}$. The SN distances in this paper confirm that the dust parameters used in \citet{yosh14} are reasonable.

\section{H$\beta$ Distances of AGNs}\label{sec:HB}

\begin{figure}
\includegraphics[scale=0.8]{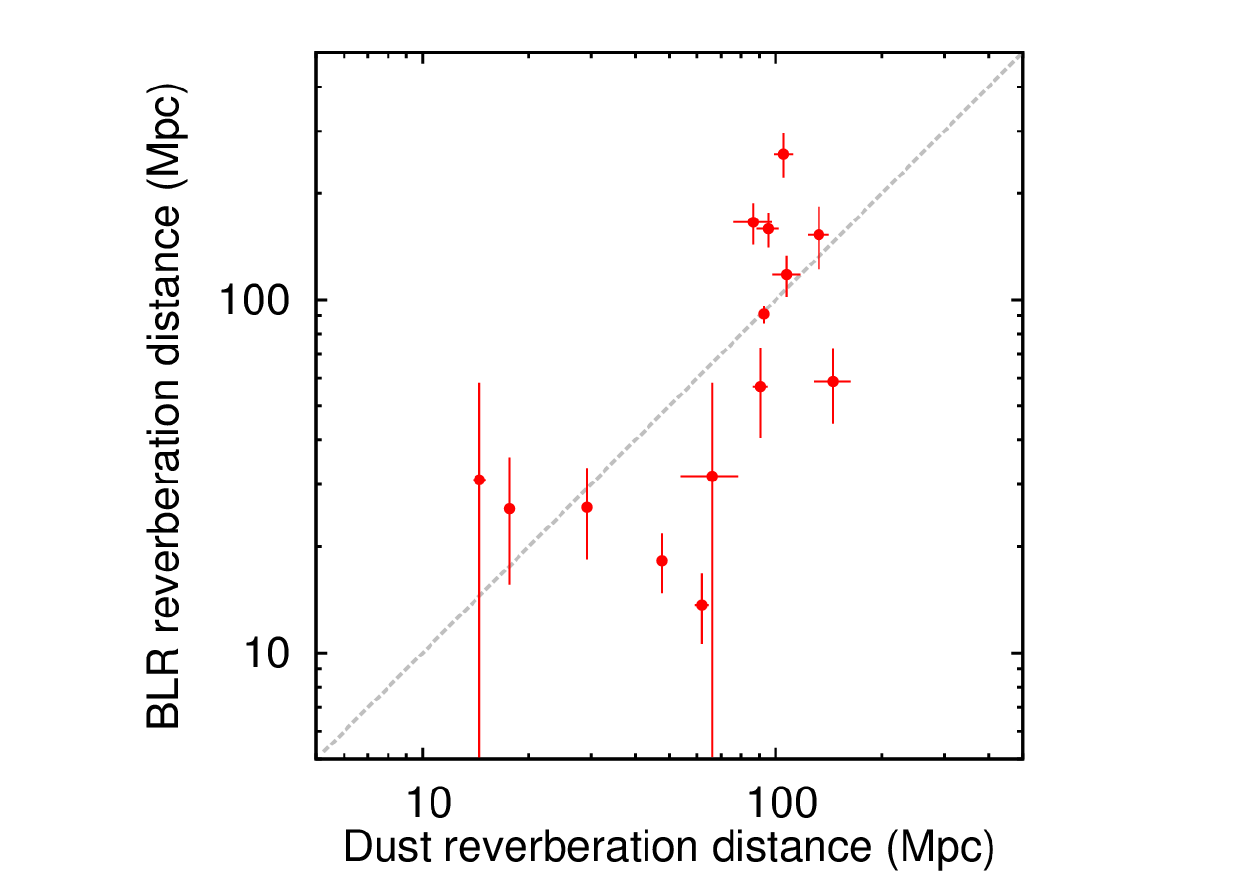}
\caption{The distances calculated using $\Delta t_{{\rm H}\beta}$ and $g_{{\rm H}\beta}=13.56$ are plotted against the distances using $\Delta t_{\rm DUST}$ and $g_{\rm DUST}=10.60$ for the 14 Seyfert 1 galaxies in our AGN monitoring program in common with \citet{wats11}.}
\label{fig:fig2}
\end{figure}

\citet{wats11} reported a relation between the relative distances for AGN based on the correlation of the BLR size with absolute luminosity. For the 14 AGNs in \citet{yosh14} that we have in common, substituting their dust reverberation distances of $d_{\rm DUST}$ into the LHS of Equation (\ref{eq1}), we use the time delay $\Delta t_{{\rm H}\beta} $ between a continuum flare and the increase in the flux in the H$\beta$ emission line from \citet{bent09} to obtain the distribution of $g_{{\rm H}\beta}$. The mean of $g_{{\rm H}\beta}$ is 13.56, and the deviation is as small as 0.08. Thus, we attribute most of the scatter to cosmic variation in the properties of the BLR.  In Figure \ref{fig:fig2} we plot their reverberation distances obtained from Equation (\ref{eq1}) using $\Delta t_{{\rm H}\beta}$ and $g_{{\rm H}\beta}=13.56$ versus $\Delta t_{\rm DUST}$ and $g_{\rm DUST}=10.60$.   

\begin{figure}
\includegraphics[scale=1.0]{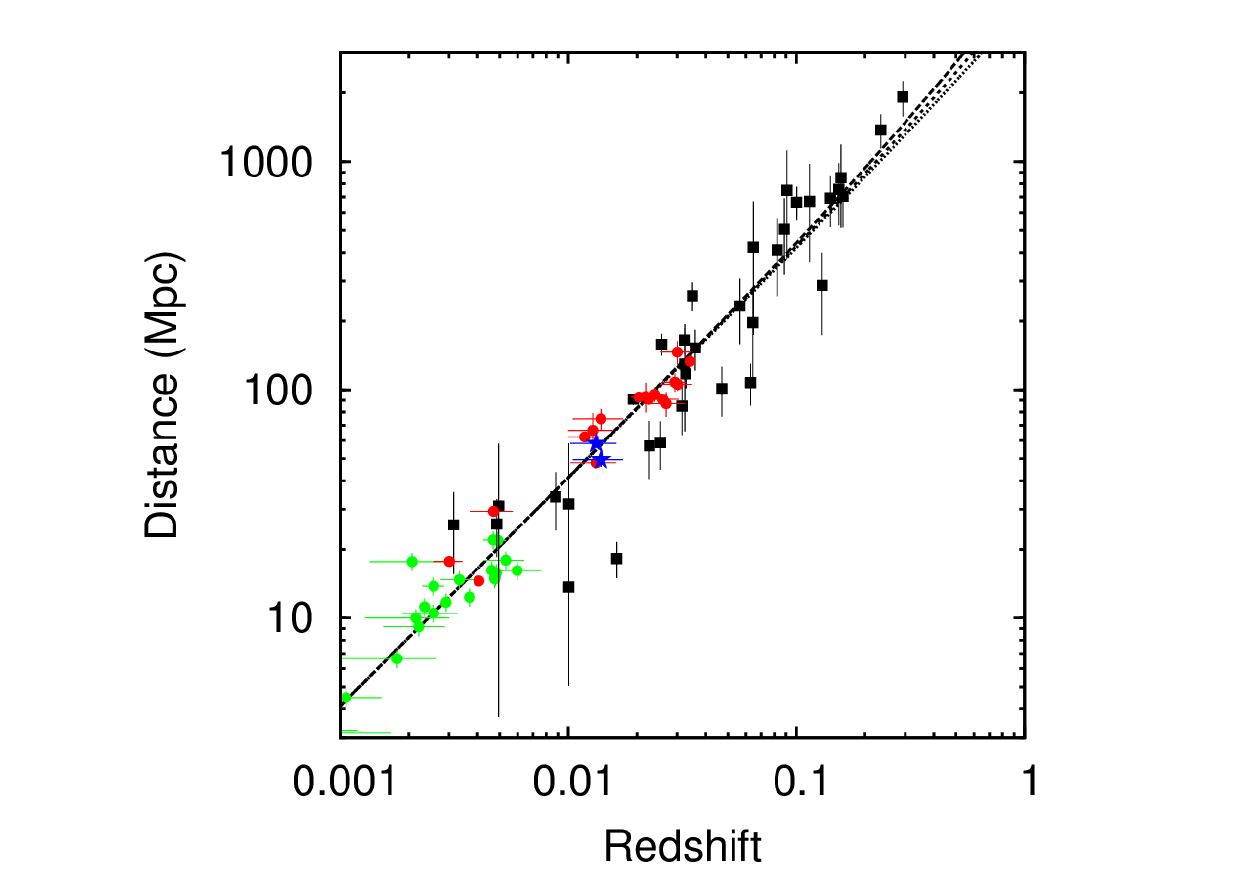}
\caption{Hubble diagram where luminosity distances are plotted against 
redshift. Black filled squares represent all the AGNs with measured 
$\Delta t_{{\rm H}\beta}$ from \citet{bent09}, and their distances 
were calculated using Equation (\ref{eq1}) with 
$g_{{\rm H}\beta}=13.56$ (see \S\ref{sec:HB}). Red filled circles 
represent AGNs with measured $\Delta t_{\rm DUST}$, and their 
distances were calculated using Equation (\ref{eq1}) with 
$g_{\rm DUST}=10.60$ \citep[see ][]{yosh14}. 
Green filled circles 
represent the galaxies with Cepheid distances
and illustrate the consistency of Cepheid and reverberation distances.
 \citep[see][]{yosh14,free01,free02}.
Blue stellate symbols represent two SNe Ia with the distances 
measured in this paper. From top to bottom, the dashed, the dot-dash, 
and the dotted lines represent the cosmological models with $(\Omega_{m}, \Omega_{\Lambda}) = (0.3,0.7)$, $(0.3, 0.0)$, and $(1.0, 0.0)$, respectively, with $H_0=73$ km s$^{-1}$ Mpc$^{-1}$. The redshift errors come from the range of model esimates of local velocity field correction.
}
\label{fig:fig3}
\end{figure}

Having determined the mean value of $g_{{\rm H}\beta}=13.56$, we use Equation (\ref{eq1}) to obtain the luminosity distances for all the AGNs with measured $\Delta t_{{\rm H}\beta}$ \citep{bent09}. These are plotted against redshift as black squares in Figure \ref{fig:fig3} along with our AGNs from \citet{yosh14} (red circles) and the galaxies with Cepheid distances \citep[][green filled circles]{free01,free02}.  

\section{Conclusion}\label{sec:CON}

The calculation in \citet{yosh14} of the calibration factor, $g=10.60$, in Equation (\ref{eq1}),
that characterizes the dust sublimation by the radiation field of the AGN central engine is 
independently supported by the distances of two SNe Ia which occurred in two different AGN host galaxies during our AGN monitoring program. 

The agreement between $g_{\rm SN}$ and $g_{\rm DUST}$ as well as the small  estimated scatter of $g_{\rm DUST}$, after excluding measurement errors (\S\ref{sec:DD}),  indicates that our characterization of the AGN dust is essentially correct and that the dust properties of all AGNs are very similar to our characterization. 

We have used 14 of our AGNs to evaluate $g_{{\rm H}\beta}$, and we have used this calibration to produce a Hubble diagram reaching $z=0.3$, that begins to distinguish between possible cosmologies  (\S\ref{sec:HB}) . With further ground based H$\beta$ observations, this could be extended to $z \approx 1$. 

\acknowledgments

We thank the staff at the Haleakala Observatories for their help with facility maintenance. This research has been supported partly by the Grants-in-Aid of Scientific Research (10041110, 10304014, 11740120, 12640233, 14047206, 14253001, 14540223, and 16740106) and the COE Research (07CE2002) of the Ministry of Education, Science, Culture and Sports of Japan.

\clearpage

\clearpage

\clearpage

\clearpage

\begin{deluxetable}{lllllll}
\tabletypesize{\scriptsize}
\tablecolumns{6}
\tablecaption{List of Target Supernovae}
\tablehead{
\colhead{\shortstack{SN\\\phn}} & 
\colhead{\shortstack{Host\\\phn}} &
\colhead{\shortstack{R.A.\\\phn}} &
\colhead{\shortstack{Dec.\\\phn}} &
\colhead{\shortstack{$z$\tablenotemark{a}\\\phn}} &
\colhead{\shortstack{$d_{\rm DUST}$\tablenotemark{b}\\(Mpc)}} &
\colhead{\shortstack{Reference\\\phn}}
}
\startdata
SN 2004bd & NGC 3786 & $11~39~42.2$ & $+31~54~31.8$ & $0.00893$ & $74.3\pm 7.8$ & IAUC 8316, 8317 \\ 
SN 2008ec & NGC 7469 & $23~03~16.6$ & $+08~52~19.8$ & $0.01632$ & $47.7\pm 1.5$ & CBET 1437, 1438
\enddata
\tablecomments{Listed references are the first report of the SN detection.}
\tablenotetext{a}{The heliocentric redshift from the NASA/IPAC Extragalactic Database (NED).}
\tablenotetext{b}{The luminosity distance of host galaxy based on the dust
reverberation of AGN \citep{yosh14}.}
\label{tab:tab1}
\end{deluxetable}%

\begin{deluxetable}{lllcll}
\tabletypesize{\scriptsize}
\tablewidth{0pt}
\tablecolumns{6}
\tablecaption{Results from Light Curve Modeling}
\tablehead{
\colhead{} & \multicolumn{2}{c}{SN 2004bd} & & \multicolumn{2}{c}{SN 2008ec} \\\cline{2-3}\cline{5-6}
  & \colhead{MLCS2k2} & \colhead{SALT2} & & \colhead{MLCS2k2} & \colhead{SALT2}
}
\startdata
  $\mu$............................. & $33.53\pm 0.24$ & $33.43\pm 0.19$ & & $33.88\pm 0.13$ & $33.79\pm 0.08$ \\
  $t_{\text{peak},B}$ (MJD)........ & $53090.31\pm 2.59$ & $53095.20\pm 1.41$ & & $54674.00\pm 0.52$ & $54674.46\pm 0.11$ \\
  $A_{V}$(mag)................ & $0.93\pm 0.29$ & $0.82\pm 0.13$\tablenotemark{a} & & $1.03\pm 0.09$ & $0.65\pm 0.07$\tablenotemark{a} \\
  Width parameter\tablenotemark{b}.. & $0.07\pm 0.12$ & $-0.15\pm 0.03$ & & $-0.19\pm 0.09$ & $-0.07\pm 0.02$ \\
  reduced $\chi^{2}$.............. & $0.519004$ & $0.072918$ & & $0.247278$ & $0.773540$
\enddata
\tablenotetext{a}{The extinction in the $V$ band is estimated from the $B$ band extinction which SALT2 calculated, adopting  $A_B/A_V =1.34$ \citep{card89}}
\tablenotetext{b}{Parameters for light curve width of SNIa. $\Delta$ in MLCS2k2 and $x_{1}$ in SALT2.}
\label{tab:tab2}
\end{deluxetable}%

\end{document}